\documentclass[10pt,pra,twocolumn,notitlepage,showpacs]{revtex4-1}

\newcommand{\itwo}{{\frac1{\sqrt{2}}}}
\newcommand\Id{\leavevmode\hbox{\small1\normalsize\kern-.33em1}}

\newcommand{\ket}[1]{\left\vert{#1}\right\rangle}

\newcommand{\ku}{\vert{0}\rangle}

\newcommand{\ave}[1]{\left\langle #1\right\rangle}

\newcommand{\ham}{{\mathcal{H}}}

\newcommand{\blist}[1]{
 \begin{list}{#1}
  { \setlength{\itemsep}{3pt}
     \setlength{\parsep}{2pt}
     \setlength{\topsep}{3pt}
     \setlength{\partopsep}{0pt}
     \setlength{\leftmargin}{1em}
     \setlength{\labelwidth}{1em}
     \setlength{\labelsep}{0.5em} } }
\newcommand{\elist}{
  \end{list}  }

\DeclareMathSymbol{\vartheta}{\mathalpha}{letters}{"12}
\DeclareMathSymbol{\theta}{\mathalpha}{letters}{"23}
\DeclareMathSymbol{\phi}{\mathalpha}{letters}{"27}
\DeclareMathSymbol{\varphi}{\mathalpha}{letters}{"1E}

\usepackage{amsmath}
\usepackage{amssymb}
\usepackage{bm}
\usepackage[usenames,dvipsnames]{color} 
\usepackage{cancel}
\usepackage{dcolumn}
\usepackage{epsf}
\usepackage{epsfig}
\usepackage{graphicx}
\usepackage{mathtools}
\usepackage{multirow}
\usepackage[final]{pdfpages}
\usepackage{setspace}
\usepackage{ulem} 
\usepackage{wrapfig}

\usepackage[usenames,dvipsnames]{color} 

\definecolor{LinkColor}{rgb}{0,0,.5}
\usepackage{hyperref}
\renewcommand{\emph}{\textit}
\graphicspath{{/Users/paola/Documents/Work/figures/Gyro/Arxiv/},{/Users/paola/Documents/Work/figures/???/}}
\begin{document}
\title {Stable  Three-Axis Nuclear Spin  Gyroscope in Diamond}
\author{Ashok Ajoy}
\author{Paola Cappellaro}
\affiliation{Department of Nuclear Science and Engineering and Research Laboratory of Electronics,
Massachusetts Institute of Technology, Cambridge, MA, USA}
\email{pcappell@mit.edu}
\begin{abstract}
We propose a sensitive and stable three-axis  gyroscope in
diamond. We achieve high sensitivity by exploiting the long coherence time of the $^{14}N$ nuclear spin associated with the Nitrogen-Vacancy center in diamond, and the efficient polarization and measurement of its electronic spin. While the gyroscope is based on a simple Ramsey interferometry scheme, we use coherent control of the quantum sensor to improve its coherence time as well as its robustness against long-time drifts, thus achieving  a very robust device with a resolution of 0.5mdeg/s/$\sqrt{\text{Hz}\cdot\text{mm}^3}$. In addition, we exploit the four axes of delocalization of the Nitrogen-Vacancy center to measure not only the rate of rotation, but also its direction, thus obtaining a compact three-axis gyroscope.
\end{abstract}
\maketitle

Gyroscopes -- sensitive rotation detectors -- find wide application in everyday life, from GPS and inertial sensing, to jerk sensors in hand-held devices and automobiles. 
Conventional gyroscopes are built using micro electro-mechanical system (MEMS) technology that allows for high sensitivities exceeding 3mdeg/s/$\sqrt{\textrm{Hz}}$ in a hundreds of micron size footprint~\cite{GyroSpec,Bhave03,Johari08}.
Despite  several advantages --including  extremely low current drives ($\sim100\mu$A) and  large bandwidths ($\gtrsim200$deg/s)-- that have allowed MEMS
gyroscopes to have ubiquitous usage, they suffer from one important drawback:
the measurement sensitivity drifts after about a few minutes of operation,
making them unattractive for long-time stable rotation sensing required for geodetic applications~\cite{Jekeli00}. 
The intrinsic reason for these drifts -- the formation of charged asperities near the surface
of the capacitive transduction mechanism -- is endemic to MEMS, but does not
occur in other  systems, such as atom interferometers~\cite{Lenef97,Gustavson97,Durfee06} or nuclear spins~\cite{Kitching11,Kornack05,Havel05}, which are used as gyroscopes. However, to achieve  sensitivities comparable to
MEMS, these systems  require large sensor volumes ($\sim\!\text{cm}^3$), long startup and averaging times,  and  large power and space overheads for excitation (lasers and magnetic fields) and detection. 

Here we propose to overcome the challenges of these two classes of gyroscope by using a 
\textit{solid-state} spin gyroscope associated with the nuclear spin of nitrogen-vacancy (NV) centers in
diamond.  Such a sensor (that we call \textit{nNV-gyro}) can achieve sensitivity on the order of $\eta\sim0.5\text{mdeg/s}/\sqrt{\text{Hz}\cdot\text{mm}^3}$, comparable to MEMS gyroscopes, while offering enhanced stability, a small footprint and fast startup-up time.  Thanks to the four possible orientations of the NV center, the nNV-gyro can operate as a three-axis gyroscope.
Furthermore, a combinatorial gyroscope, consisting of a MEMS system coupled to the nNV-gyro  may allow for the best of both worlds --
rotational sensing with ultra-high sensitivity and long-term stability.

The nNV-gyro combines the efficient initialization and measurement offered by the NV electronic spin, thanks to its optical polarization and fluorescent readout, with the stability and long coherence time of the nuclear spin, which is preserved even at high density. 

\begin{figure}
\centering
\includegraphics[scale=0.4]{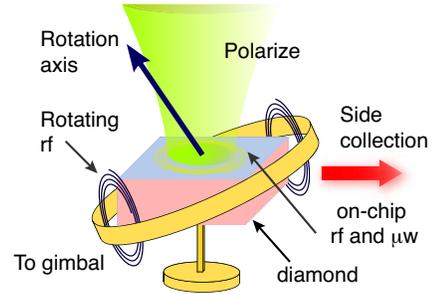}
\caption{A schematic design of the nNV-gyro. A slab of diamond of $2.5\times2.5$mm$^2\times150\mu$m is anchored to the device body. RF coils and microwave coplanar waveguides are fabricated on the diamond for fast control. The NV centers are polarized by a green laser (532nm) and state-dependent fluorescence intensity (637nm) is collected employing a side-collection technique~\cite{Lesage12}.  A second set of rf coils rotates with respect to the diamond chip frame, for example by being attached to one or more rings mounted in a mechanical gimbal gyroscope configuration (not shown).  
}
\label{fig:setup}
\end{figure}
\begin{figure}[t]
\centering
\includegraphics[scale=0.18]{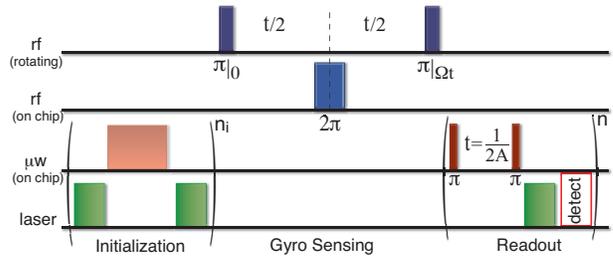}
\caption{nNV-gyro control sequence. rf pulses, resonant with the quadrupolar transition of the $^{14}$N nuclear spin, are applied in a frame rotating at a rate $\Omega$ with respect to the diamond, thus inducing a phase $\Omega t$. The nuclear spin is first initialized by polarization transfer from the NV electronic spin~\cite{SOM}. An echo pulse is applied --in the diamond frame-- to refocus static frequency shifts. Finally, $\Omega$ is extracted by mapping the nuclear spin phase shift onto a population difference of the electronic spin, and measuring the corresponding fluorescence intensity. }
\label{fig:sequence}
\end{figure}

Similar to NMR gyroscopes, the operating principles are based on the detection of the  dynamical phase~\footnote{While this manuscript was in preparation, a proposal to utilize the \textit{geometric} phase acquired by the electronic NV spin as a rotation sensor has appeared in \cite{Ledbetter12x,Maclaurin12x}} that the Nitrogen-14 nuclear spin-1 acquires when it rotates around its symmetry axis.  Consider an isolated spin 1 in a diamond crystal, with  Hamiltonian $\ham_0=QS_z^2+\gamma_NbS_z$, where $b$ is a small magnetic field, $\gamma_Nb\ll Q$. The spin is  subject to radio-frequency (rf) fields in the transverse plane at the frequency $Q$, on resonance with the $0\leftrightarrow\pm1$ transition. The diamond rotates around the spin symmetry axis (z-axis) at a rate $\Omega$. In a  Ramsey sequence (see Fig.~\ref{fig:sequence}) the spin will  acquire a phase $\phi=(\gamma_Nb+\Omega)t$, as the phase of the applied pulses changes at a rate $\Omega$ in the frame co-rotating  with the diamond. The  sensitivity $\eta$  to the rotation rate is shot-noise-limited  and for unit measurement time is ideally given by $\eta\propto1/\sqrt{tN}$, where $N$ is the number of nitrogen nuclear spins associated with NV centers in the diamond chip.
The performance of the nNV-gyro  is then set by the coherence time of the spin, which together with the efficiency of its initialization and readout sets the limits to the sensitivity, and by its robustness to external parameters, such as temperature, stray fields and strains, which determines its stability. In the following we analyze  these effects while presenting  details of the nNV-gyro operation.

The nuclear spin is first initialized by polarization transfer from the electronic NV spin. The combination of a cycling transition on the electronic $m_s=0$ levels and a transition through a metastable level for the $m_s=\pm1$ levels, yields high polarization of the electronic spin~\cite{Robledo11b,Waldherr11b}.  The polarization can be transferred to the nuclear spin by various methods, including adiabatic passage~\cite{Fuchs11},  measurement post-selection~\cite{Waldherr12} or exploiting a level anti-crossing in the orbital excited state at  $\sim 500$G magnetic field~\cite{Smeltzer09,Jacques09,Fischer12x}. It is as well possible to induce polarization transfer at low (or zero field) either in the rotating frame~\cite{Luy00} or by exploiting two-photon transitions via longitudinal driving~\cite{Childress10}. Working at low magnetic field simplifies the nNV-gyro operation and  avoiding the level anti-crossing   allows repeated readouts~\cite{Neumann10b}, while still allowing high fidelity preparation in a time $t_{pol}\leq2\mu$s (see online material for details). A small bias field $B\sim20$G can  be applied to separate the signal from the four different NV classes, associated with the four $\ave{111}$ directions in diamond, by shifting them off-resonance. We note that the initialization time is much shorter than the startup required even for MEMS gyroscopes (a few tens of milliseconds).

For ease of operation, we assume that the rf and microwave ($\mu$w) pulses used for initialization and readout can be delivered by an on-chip circuit, integrated with the diamond chip. After preparation, the NV electronic spin is left in the $\ku$ state, which does not couple to the $^{14}$N nuclear spin nor to the spin bath. Then a Ramsey sequence is applied using the off-chip rf driving. A $2\pi$-pulse at the center of the sequence, applied with the on-chip rf refocuses the effects of stray magnetic fields and provides decoupling from the spin bath. We note that the echo will be effective even if the on- and off-circuit cannot be made phase-coherent. With this control sequence, the coherence time of the sensor spin is thus limited by $T_2$ (and not by the much shorter dephasing time $T_2^*$), which can be exceptionally long for nuclear spins~\cite{Fuchs11,Kolkowitz12}. It is then possible to operate at very high density of sensor spins. Even at a NV density  $n_{NV}\sim10^{18}$cm$^{-3}$ and assuming a density of single nitrogen defects (P1 centers~\cite{Hanson08,Acosta09,Wee07}) $n_{P1}\approx10n_{NV}\sim10^{19}$cm$^{-3}$, the $^{14}$N $T_2$ time is not appreciably affected by the P1 bath. Indeed, while the dipole-dipole interaction among P1 centers is very large at these densities ($\sim3$MHz), the coupling to the nuclear spin is still small ($\sim 345$Hz) due to its low gyromagnetic ratio. This leads to motional narrowing and a very slow exponential decay,   
 as confirmed by simulations based on a cluster expansion approach~\cite{SOM,Maze08,Witzel08}. 
 The $^{14}$N coherence time is also affected by the interaction with the close-by NV center~\cite{Fuchs11,Waldherr12}, which induces dephasing when undergoing relaxation with $T_1\sim2-6$ms  at room temperature and low field~\cite{Jarmola12x}. While in isotopically purified, low defect density diamonds at high magnetic field the \textit{dephasing} time $T_2^*$ can be as long 7ms~\cite{Waldherr12}, in the proposed conditions of operation we can  conservatively estimate the \textit{coherence} time of the nuclear spin to be $T_2=1$ms.   We note that the echo sequence has the added benefit to make the measurement insensitive to many other imperfections, such as the temperature variation, strain, background stray fields, variation in the quadrupolar interaction, instability in the applied bias magnetic field,  etc. Thus this scheme yields a very robust and stable gyroscope.

After the sensing sequence, the $^{14}$N spin is left in the state 
\begin{equation}
\ket{\psi_n}=\itwo\sin(\Omega t)\left(e^{-i\Omega t}\ket{-1}-e^{i\Omega t}\ket{+1}\right)-\cos(\Omega t)\ku,
\end{equation} 
which can be mapped  into a population difference between the NV levels, thanks to the $A\approx2.2$MHz hyperfine coupling. The time required to map the state onto the NV center is $t_{map}=230$ns, which is close to the $T_2^*$ time for the NV, thus we expect a reduction in contrast due to the NV dephasing. Indeed it is the NV dephasing time that ultimately limits the allowed spin densities. A possible solution would be  to perform a spin echo on both nuclear and electronic spins to extend the coherence time. 
The nuclear-electronic spin system is described by the Hamiltonian
\begin{equation}\ham_{ne}=\Delta S_z^2+\gamma_ebS_z+AS_zI_z+QI_z^2+\gamma_n I_z,\end{equation}
where we only retained the longitudinal component of the isotropic hyperfine interaction because of the large zero-field splitting of the electronic spin, $\Delta=2.87$GHz. The readout sequence (Fig.~\ref{fig:sequence}), with pulses on resonance to both $0\leftrightarrow\pm1$ transitions generate the state~\footnote{the mapping is also possible if only one NV transition is driven, although it requires a longer time.} 
\begin{equation}\begin{array}{l}\ket{\psi_{ne}}=-\cos(\Omega t)\ket{00}+\sin(\Omega t)\times\\\qquad\left[e^{i\Omega t}(\ket{\text{-}1\text{-}1}+\ket{\text{+}1\text{-}1})+e^{-i\Omega t}(\ket{\text{+}1\text{+}1}+\ket{\text{-}1\text{+}1})\right]/2\end{array}\end{equation}
Optical readout can then extract the information about the rotation $\Omega$. We note that the measurement step can be repeated to improve the contrast~\cite{Jiang09,Neumann10b}: although at low field the effective relaxation time of the nuclear spin under optical illumination is short, thus limiting the number of repeated readouts~\cite{Jiang09}, when combined with high collection efficiency $\eta_m\approx 1$~\cite{Lesage12} we can still achieve a detection efficiency $C\sim0.25$ 
 for $n_r=100$ repetitions and a total readout time $t_{ro}\approx 150\mu$s~\cite{SOM}. 
 The higher detection efficiency will also allow to achieve a large dynamic-range by exploiting adaptive phase estimation schemes~\cite{Cappellaro12,Waldherr12,Nusran12}. 
 \begin{figure}
\centering
\includegraphics[scale=0.5]{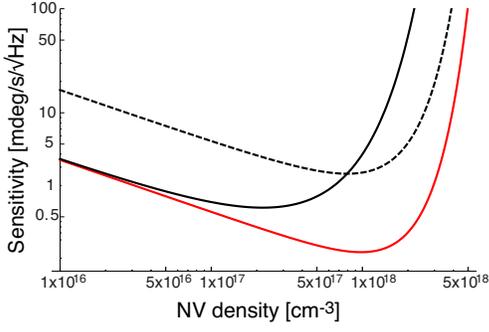}
\caption{nNV-gyro sensitivity in mdeg/s/$\sqrt{\text{Hz}}$, as a function of density. We considered a diamond chip of size $(2.5\times2.5)$mm${^2}\times150\mu$m and an interrogation time $t=1$ms (solid lines) and $t=0.1$ms (dashed line). The sensitivity for a simple Ramsey scheme (black lines) is limited by the nuclear spin $T_2^*$. Using an echo scheme (red thick line) improves the sensitivity, which is now limited by the coherence time of the NV electronic spin used to read out the nuclear spin state.}
\label{fig:eta}
\end{figure}
The sensitivity can thus be estimated taking into consideration all the parameters and inefficiencies presented:
\begin{equation}\eta=\frac{\sqrt{T_2+t_d}}{ CT_2\sqrt{N}},\end{equation}
where we introduce the dead-time $t_d=t_{ro}+t_{pol}$, including the initialization and readout  time. For a volume of $V=1$mm$^3$ and $N=n_{NV}V/4\approx2.5\times10^{14}$ sensor spins, the estimated sensitivity for the nNV-gyro is thus $\eta\approx 0.5 ($mdeg/s$)/\sqrt{\text{Hz}}$, 
better than current MEMS gyroscopes (see Fig. \ref{fig:eta}).  The stability of the nNV-gyro can however be much higher and comparable to atomic gyroscopes. Indeed the echo-based scheme makes the nNV-gyro insensitive to long-time drifts due to temperature and stray fields. In addition, the NV spin is a sensitive probe of these effects, capable of measuring magnetic~\cite{Taylor08} and electric~\cite{Dolde11} fields, as well as frequency~\cite{Hodges11} and temperature shifts~\cite{Acosta10,Toyli12}. The NV spin could then be used to monitor such drifts and correct them via a feedback mechanism.

\begin{figure}[b]
\centering
\includegraphics[scale=0.4]{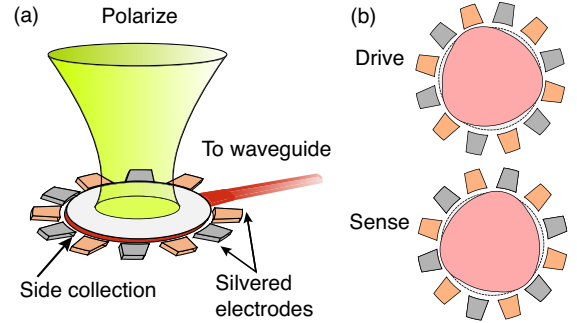}
\caption{
An integrated nNV-MEMS gyroscope, comprising a 
bulk acoustic wave (BAW) single-axis MEMS disk gyroscope~\cite{Johari08}  in a
diamond substrate implanted with NV centers. The disk  acts as a mechanical gyroscope, while the nuclear spins contained in it form a spin gyroscope.  (a) Schematic  of the integrated gyroscope operation.  The spins implanted in the disk are polarized by an on-chip green laser. 
The electrodes surrounding the disk are silvered to allow for total internal reflection and fluorescence is collected by replacing one of the electrodes by an optical waveguide at 638nm.  Striplines for rf/$\mu$w control are fabricated on the disk. 
(b) Operation of the BAW mechanical gyroscope. The BAW is electrostatically driven in the second elliptic mode through
the drive (orange) electrodes. A rotation out of the plane causes a decrease in the gap near the sense (grey) electrodes
leading to a capacitive measurement of the rotation. Combinatoric filtering with the nNV measurement would lead to noise rejection and improved stability.}
\label{fig:MEMS}
\end{figure}
Until now, we assumed that the relative rotation of the rf pulses with respect to the diamond was applied along the NV axis of symmetry. The nNV-gyro can however work as a three-axis gyroscope, extracting information about the rotation rate as well as its direction. If the rotation axis is at an angle $\{\theta,\phi\}$ with respect to the NV axis, the $^{14}$N undergoes a complex evolution since the second rf pulses is not only rotated by an angle $\psi(\theta,\phi,\Omega t)$ in the NV x-y plane, but it also has a smaller flip angle $\alpha(\theta,\phi,\Omega t)<\pi$~\cite{SOM}. The state at the end of the Ramsey sequence is then
\begin{equation}\begin{array}{l}\ket{\psi_n}=\frac{e^{-i \psi } \left[\sin (\psi )-i \cos \left(\frac{\alpha }{2}\right) \cos (\psi )\right]}{\sqrt{2}}\ket{+1}\\-\sin(\alpha/2)\cos(\psi)\ku 
-\frac{e^{i \psi } \left[\sin(\psi )+i \cos\left(\frac{\alpha }{2}\right) \cos(\psi)\right]}{\sqrt{2}}\ket{-1}\end{array}
\label{eq:axes}
\end{equation}
The nuclear spins associated with the four different NV classes will therefore experience different evolutions; the signal from each class can be  measured
by sequentially mapping each class onto the corresponding electronic spin via on resonance $\mu$w pulses (a bias field of 10-20G is sufficient to lift the frequency degeneracy among the 4 classes~\cite{Maertz10,Steinert10}). We note that a more efficient scheme would take advantage of the repeated readouts and long relaxation times of the nuclear spins, to measure the signal from 3 NV classes without the need to repeat the Ramsey interrogation sequence.

In conclusion, we have proposed a compact solid-state device able to measure rotation rates with a resolution $\eta\approx0.5$mdeg/sec/$\sqrt{\text{Hz}}$, better than state of the art MEMS gyroscopes, while providing improved stability. The device takes full advantage of the long coherence time of  \textit{nuclear} spins, which is preserved even at very high densities, while exploiting its interaction with the electronic spin of the NV center for efficient initialization and readout.
Going beyond our conservative estimates, the nNV-gyro could achieve the performance of inertial-grade gyroscopes with improvements in the coherence time of the nuclear and electronic NV spin, as could be obtained with $\gtrsim50\%$ N to NV conversion efficiency~\cite{Pezzagna10}, with a preferential alignment of the NV symmetry axis along two directions~\cite{Edmonds11x}, and with improved collection efficiency, exploiting single-shot measurement of the NV center at low temperature~\cite{Robledo11,Togan11}. Alternatively, one could consider smaller devices --at the micron scale-- and exploit its long-time stability to improve the performance of MEMS gyroscope (see Fig.~\ref{fig:MEMS}). 

\textbf{Acknowledgments:} It is a pleasure to thank Sunil Bhave for stimulating discussions and encouragement. We thank  Jonathan Hodges for discussions. This work was supported in part by the
U.S. Army Research Office through a MURI grant No. W911NF-11-1-0400.

\bibliographystyle{apsrev4}
%
\newpage.\newpage

\appendix
\section{Supplementary Material}
\vspace{96pt}
\subsection{Polarization scheme for the $^{14}$N nuclear spin}
Polarization transfer between the NV electronic spin and the $^{14}$N nuclear spin is complicated by the fact that both spins are spin 1. 
Unlike for spin-1/2, polarization transfer in the rotating frame (under the Hartmann-Hahn matching condition) does not lead to perfect polarization.
Under the high-density conditions here considered, relying on selective pulses would require times longer than the coherence time of the electronic spin. Solutions to this problem have been proposed either working close to avoided crossing, where the energy of the nuclear and electronic spins are on resonance~\cite{Fuchs11,Smeltzer09,Jacques09} or by using a probabilistic, measurement-based method~\cite{Waldherr12}. Both these approaches are viable, but have drawbacks. The first method prevents the use of repeated readouts while the second method is too lengthy.
Here we propose to use forbidden two-photon transitions to drive the population transfer. Driving the NV electronic spin at the $\Delta\pm\gamma b-Q$ transition with a field \textit{along} its longitudinal (z) axis\cite{Childress10},  modulates its resonance frequency thus making energy exchange with the nuclear spin possible. This is similar to two-photon transitions allowed by Floquet theory~\cite{Shirley65}. While these transition rates are usually small, the ability to drive the NV electronic spin with very high fields~\cite{Fuchs09}, makes the polarization time $t=\pi\frac{\Delta+\gamma b+Q}{A\Omega_R}$ short: for a Rabi frequency $\Omega_R=500$MHz and a field of $20$G, the time required is only $1.3\mu$s. Given the short dephasing time of the NV spin, a two-step process might be required, in which the NV is optically re-polarized before driving the polarization exchange a second time.
\begin{figure}[th]
\centering
\includegraphics[scale=0.19]{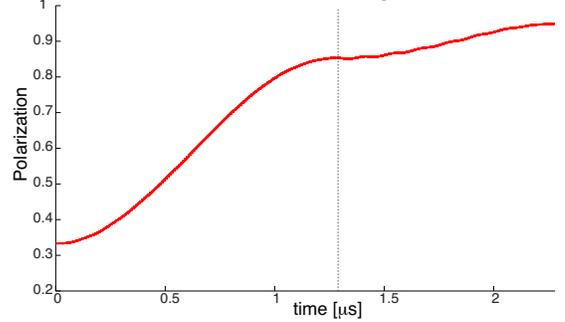}
\caption{Polarization of the $^{14}N$ nuclear spin under longitudinal drive of the NV electronic transitions, with a Rabi frequency $\Omega_R=500$MHz and a field of $20$G. The simulation includes dephasing of the electronic spin modeled by a Ornstein-Uhlenbeck process yielding a $T_2^*$ time of about $200$ns. To achieve high polarization the process is repeated twice by re-polarizing the NV electronic spin (dashed line) via optical illumination.}
\label{fig:T2sims}
\end{figure}

\subsection{Operation of the nNV-gyro as a three-axis gyroscope}
\begin{figure}[hb]
\centering
\includegraphics[scale=0.8]{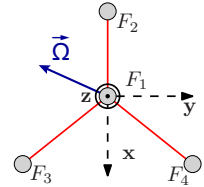}
\caption{Symmetry axes of the NV center. In an ensemble of NV centers, the spins will belong to four different families ($F_1-F_4$) depending on their symmetry axis. The rotation vector $\Omega$ will thus induce different dynamics of the four classes which can be used to reconstruct the magnitude and direction of rotation.}
\label{fig:axes}
\end{figure}

The Nitrogen-vacancy center in diamond consists of a substitutional
nitrogen adjacent to a vacancy in the lattice.  The N-to-vacancy axis sets the direction of the zero-field splitting 
axis can then  be along any of the four tetrahedral $\ave{111}$
crystallographic directions of the diamond lattice (see Fig.\ref{fig:axes}). This intrinsic symmetry can be exploited to operate the  nNV-gyro  as a three-axis gyroscope, extracting information about the rotation rate as well as its direction.

While the maximum sensitivity to the rotation rate is achieved for the axis of rotation aligned with the symmetry axis, 
if the rotation $\vec{\Omega}$ is about an axis forming an angle $\{\theta,\phi\}$ with respect to the NV axis, the $^{14}$N still undergoes a complex evolution that depends on the rotation. 
The second rf pulses is  rotated by an angle $\psi(\theta,\phi,\Omega t)$ in the NV x-y plane, with 
\begin{widetext} 
\[\arctan(\psi)=\frac{4 \left(\sin ^2(\theta ) \sin (2 \phi ) \sin ^2\left(\frac{\Omega t}{2}\right)+\cos (\theta ) \sin (\Omega t)\right)}{2 \sin ^2\left(\frac{\Omega t}{2}\right) \left(\cos (2 \phi )-2 \cos (2 \theta ) \cos ^2(\phi )\right)+3 \cos (\Omega t)+1}\]
The flip angle $\alpha(\theta,\phi,\Omega t)<\pi$ is also reduced with respect to the nominal angle, 
\[\alpha=\frac{1}{16} \left[2 \sin ^2\left(\frac{\Omega t}{2}\right) \left(\cos (2 \phi )-2 \cos (2 \theta ) \cos ^2(\phi )\right)+3 \cos (\Omega t)+1\right]^2+\left[\sin ^2(\theta ) \sin (2 \phi ) \sin ^2\left(\frac{\Omega t}{2}\right)+\cos (\theta ) \sin (\Omega t)\right]^2\]
\end{widetext} 

 The state at the end of the Ramsey sequence is then given by 
\begin{equation}\begin{array}{l}\ket{\psi_n}=\frac{e^{-i \psi } \left[\sin (\psi )-i \cos \left(\frac{\alpha }{2}\right) \cos (\psi )\right]}{\sqrt{2}}\ket{+1}\\-\sin(\alpha/2)\cos(\psi)\ku 
-\frac{e^{i \psi } \left[\sin(\psi )+i \cos\left(\frac{\alpha }{2}\right) \cos(\psi)\right]}{\sqrt{2}}\ket{-1},\end{array}
\label{eq:axes}
\end{equation}
where the angles $\alpha$, $\psi$ are different for each family of NVs. 
Measuring the signal from three families thus allows extracting information about $\vec{\Omega}$. While this scheme requires along the transverse direction of each family, using a single axis $\mu$w is also possible, although it makes the deconvolution algorithm more complicated.
  
\subsection{Coherence time of the $^{14}$N nuclear spin.}
We simulated the $^{14}$N nuclear spin evolution under a Ramsey and echo sequence. The simulation was performed using a cluster expansion to model the effects of a P1 electronic spin bath. The spin bath was modeled as an ensemble of electronic spin-1/2, thus omitting the details of the actual bath (such as the strong hyperfine coupling of the P1 to its nuclear spins, as well as shielding effects due to nearby carbon-13 nuclear spins~\cite{Bar-Gill12x} and localization effects due to disorder~\cite{Witzel12x}). Still, the simulations can give rough estimates of the expected $T_2$ times and are consistent with the results of a simple model based on describing the spin bath as a classically fluctuating bath described by a Ornstein-Uhlenbeck process. In particular, we expect that due to motional narrowing, the coherence times are long even at high densities (see Fig.~\ref{fig:T2sims}).
\begin{figure}
\centering
\includegraphics[scale=0.6]{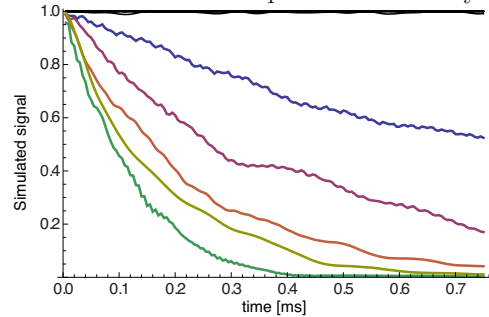}
\caption{$^{14}N$ coherence decay under a spin-echo sequence (black lines) and a Ramsey sequence. We simulated an ensemble of hundred non-interacting nuclear spin, each subjected to a bath of $\sim 500$ electronic spins, with a density (from bottom to top lines) of $n=(0.35, 1.06, 1.76,2.47,3.17)\times 10^{19}$cm$^{-3}$ (The lines for the echo are indistinguishable).}
\label{fig:T2sims}
\end{figure}
\subsection{Improved measurement efficiency by repeated readouts}
The detection efficiency is given by $C=\left(1+\frac{2(n_0+n_1)}{(n_0-n_1)^2}\right)^{-1/2}$~\cite{Taylor08}, where $n_{1,0}$ is the number of photons collected if the NV spin is in the $m_s=\{0,1\}$ state, respectively. In the repeated readout scheme~\cite{Jiang09,Neumann10b}, the state of the nuclear spin is repetitively mapped onto the electronic spin, which is then read out under laser illumination. The measurement projects the nuclear spin state into a mixed state, but the information about its population difference is preserved, under the assumption that the measurement is a good QND (quantum non-demolition) measurement. We can include the results of these repeated readout by defining a new detection efficiency, $C_{n_r}=\left(1+\frac{2(n_0+n_1)}{n_r(n_0-n_1)^2}\right)^{-1/2}$, which shows an improvement $\propto\sqrt{n_r}$, where $n_r$ is the number of measurement. The sensitivity needs of course to be further modified to take into account the increased measurement time. Provided the time needed for one measurement step is smaller than the interrogation time (including the initialization time), it becomes advantageous to use repeated readouts. The maximum number of readouts is set by the nuclear spin relaxation under optical illumination, driven by non-energy conserving flip-flops in the excited states.  While at high fields, this time is very long, allowing 2000 measurements in 5ms, in the present conditions we find that $\sim100$ measurements would provide a good balance between the improvement in $C$ and the reduction due to longer measurement time.

\bibliographystyle{apsrev4P}

%
\end{document}